\newcommand {\pT}       {\ensuremath{p_{\mathrm T}}}

\documentclass[12pt]{iopart}

\usepackage{iopams}  
\usepackage{graphicx}
\usepackage{subfig}
\usepackage[countmax]{subfloat}

\begin{document}

\title[Jet Reconstruction with Particle Flow]{Jet Reconstruction with Particle Flow in Heavy-Ion Collisions with CMS}

\author{Matthew Nguyen for the CMS Collaboration}

\address{CERN}
\ead{Matthew.Nguyen@cern.ch}
\begin{abstract}

In the particle-flow approach information from all available sub-detector systems is combined to reconstruct all stable particles.  The global event reconstruction has been shown to improve, in particular, the resolution of jet energy and missing transverse energy in $pp$ collisions compared to purely calorimetric measurements. This improvement is achieved primarily by combining the precise momentum determination of charged hadrons in the silicon tracker with the associated energy depositions in the calorimeters. By resolving individual particles inside jets, particle flow reduces the sensitivity of the jet energy scale to the jet fragmentation pattern, which is known to be one of the largest sources of systematic uncertainty in jet reconstruction. Particle flow reconstruction is thus potentially well-suited for the study of potential modifications to jet fragmentation in heavy-ion collisions. The particle flow algorithm has been adapted to the heavy-ion environment. The performance of jet reconstruction from particle flow objects in PbPb collisions using the anti-$k_{\rm T}$ jet reconstruction algorithm is presented.

\end{abstract}


The CMS detector~\cite{cms} contains four primary sub-detector systems, which are, in order of increasing distance from the beam, a high precision silicon tracking system, a high resolution, high granularity electromagnetic calorimeter (ECAL), a somewhat coarser hadron calorimeter (HCAL), and finally muon chambers.  Using a nearly 4 $T$ magnetic field, this combination of sub-detectors, allows one, in principle, to distinguish five categories of stable or pseudo-stable particles:  charged hadrons, photons, neutral hadrons, muons and electrons. 

In the approach traditionally used at hadron colliders, jets are reconstructed exclusively from calorimeter information (Calo jets).  In CMS jets are clustered from calorimeter towers which consist of a single HCAL cell and the subtended 5x5 set of ECAL crystals.  As the tower dimensions are determined by the HCAL geometry, the ECAL granularity is not exploited in this approach.  The jet resolution is driven by the energy resolution of the HCAL ($\sigma_{E}/E \approx 100\%/\sqrt{E}$).  Moreover the particle \pT\ dependence of the calorimeter response is not corrected for, leading to a non-linear response to jets, causing the jet response to depend on the fragmentation pattern of jets, which translates into an increased uncertainty on the jet energy scale.  To further complicate matters, at low \pT, due to the strong magnetic field, charged hadrons are bent outside of a typical jet radius.   

An alternate approach, known as particle flow (PF), individually reconstructs all stable particles by combining information from all available sub-detectors.  These reconstructed particle may then be used as input to a jet reconstruction algorithm in place of calorimeter towers.  By making use of the high precision silicon tracker, one may improve the \pT\ resolution of charged hadrons, and largely circumvent the non-linearity of the HCAL response.  The CMS implementation of PF has been shown to largely improve the jet \pT\ and angular resolution in $pp$ collisions~\cite{PFpp}.  In heavy-ion collisions the same PF algorithm is employed.  However, charged particle tracks are reconstructed with a different algorithm to cope with the high multiplicity of PbPb collisions.  A hit is required in each of the three layers of the pixel detector which keeps the fake rate very low, even for the most central collisions, but reduces the efficiency to a level of 60-70\%, compared to about 90\% for the $pp$ algorithm.  Charged hadrons for which no track is reconstructed default to a purely calorimetric measurement (corrected for the expected hadron response), degrading the jet energy resolution compared to the $pp$ algorithm~\cite{PFpp}.

\begin{figure}[ht!]
\centering
\subfloat[]
{
\label{fig:subfig1}
\includegraphics[width=0.44\textwidth]{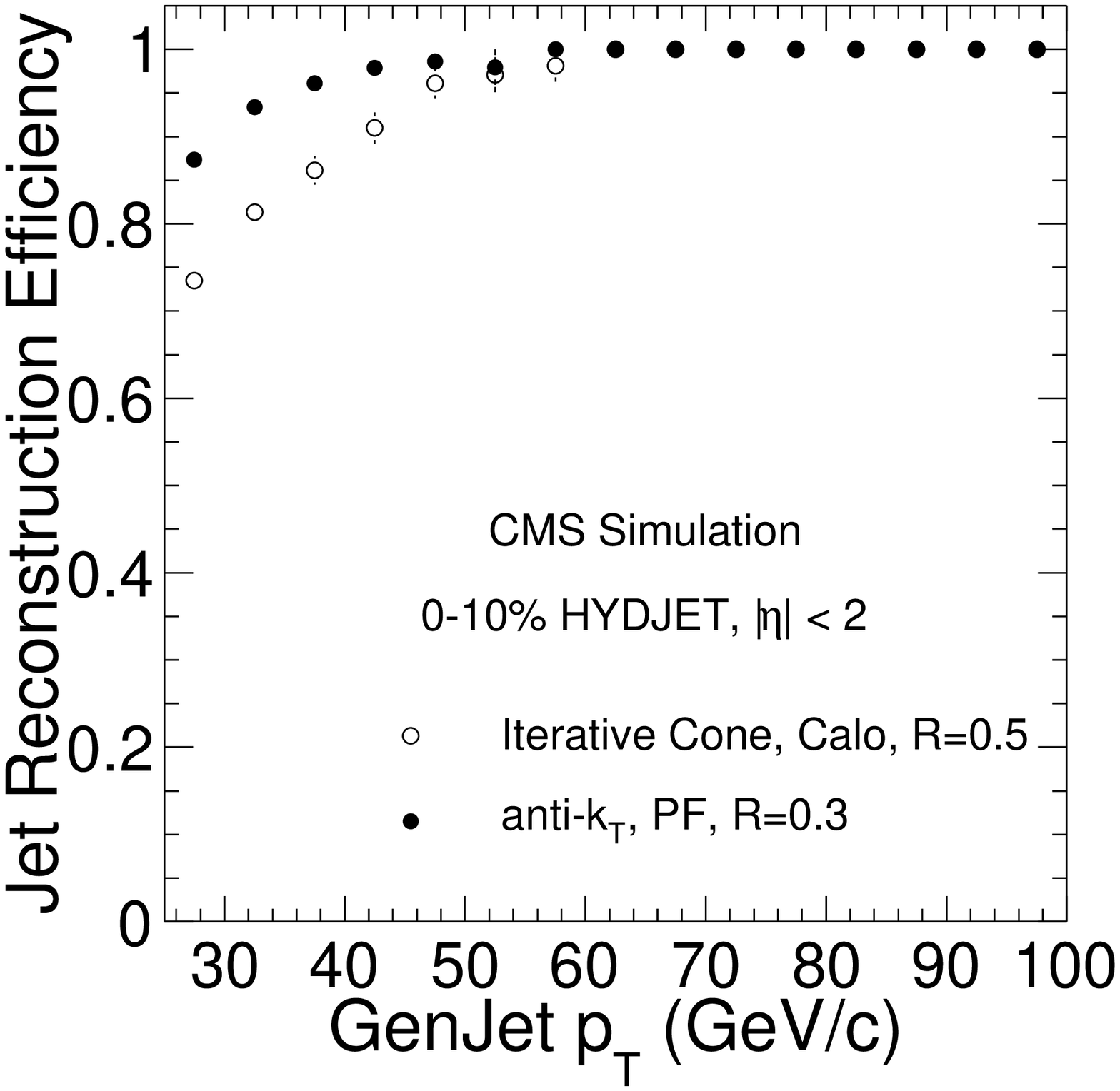}
}
\subfloat[]
{
\label{fig:subfig2}
\includegraphics[width=0.44\textwidth]{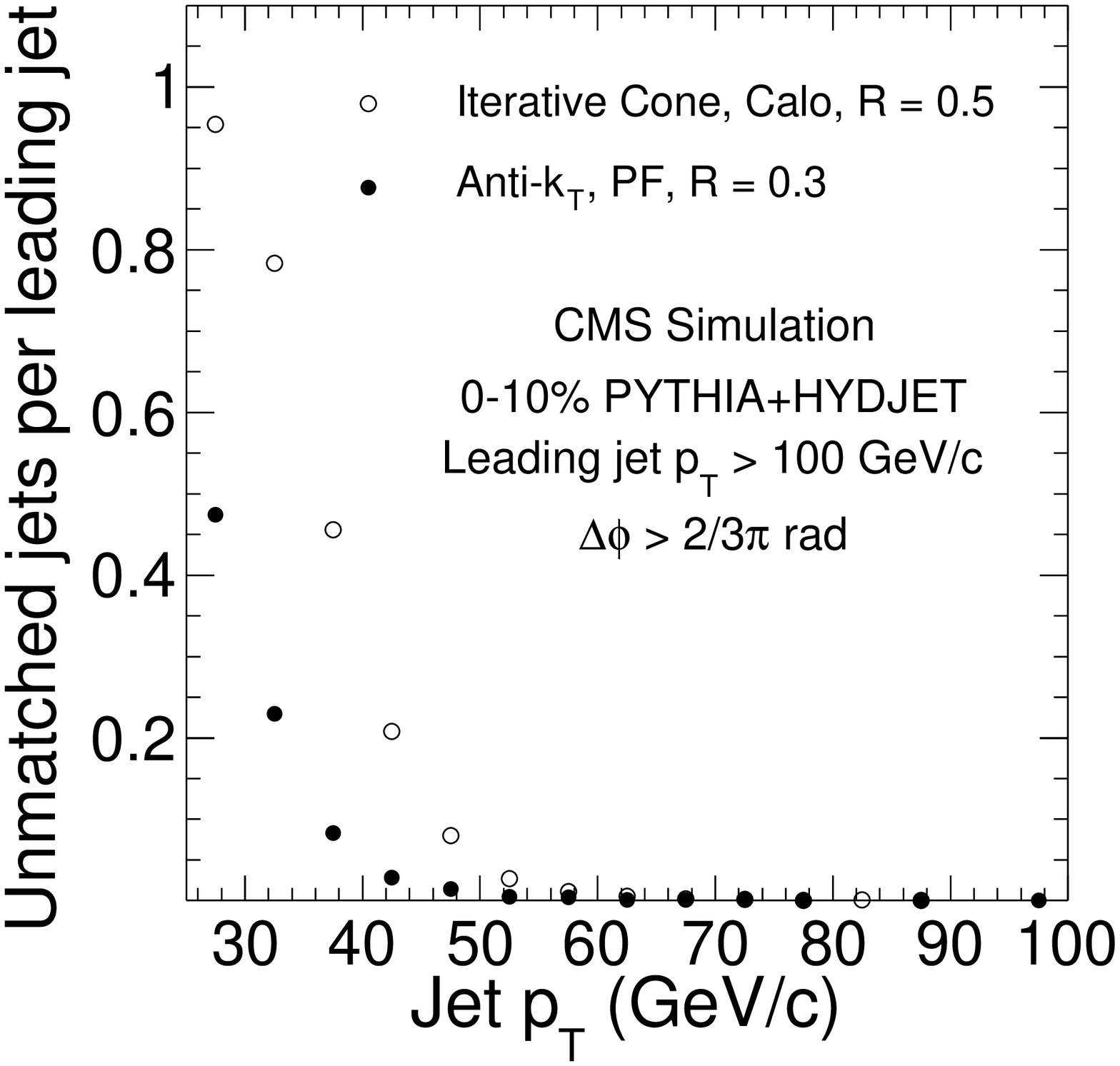}}

\label{fig:subfigureExample}
\caption[Optional caption for list of figures]{\subref{fig:subfig1} The jet reconstruction efficiency in 10\% most central {\sc{hydjet}} events.  \\ \subref{fig:subfig2} Unmatched jets per leading jet in 10\% most central {\sc{pythia+hydjet}} events.
}
\end{figure}

Jet reconstruction in PbPb collisions proceeds as follows.   First, the underlying event is subtracted according to the procedure outlined in~\cite{Kodolova} and first employed in~\cite{ourPaper}.  As is the case for $pp$ collisions, the jet energy corrections (JEC) are derived from {\sc{pythia}}.  Data-driven techniques in $pp$, such as $\gamma$+jet balancing, show that these corrections are valid to within a few \%.  The validity of this factorized background subtraction and JEC approach can then be tested in {\sc{pythia}} events embedded in PbPb events simulated with {\sc{hydjet}} ({\sc{pythia+hydjet}})~\cite{hydjet}.  Jets are reconstructed with the anti-$k_{\rm T}$ algorithm with $R$ = 0.3.  Figure~\ref{fig:subfig1} shows the jet reconstruction efficiency in 10\% most central {\sc{hydjet}} events for both PF jets and Calo jets reconstructed with a cone algorithm with $R$ = 0.5 as described in~\cite{ourPaper}.  The PF jet reconstruction is fully efficient by about 50 $GeV/c$, a somewhat lower value than for Calo jets, mostly owing to the smaller value of $R$.

For an analysis of dijets, it is important not only to find jets, but to be able to select jet pairs from the same hard scattering.  This issue is investigated in {\sc{pythia+hydjet}}.  First, a leading jet is selected with \pT\ $>$ 100 $GeV/c$.  We then count the rate at which reconstructed jets on the away-side cannot be matched in $\Delta R$ to an embedded {\sc pythia} jet.  Figure~\ref{fig:subfig2} shows the rate of unmatched jets per leading jet as a function of reconstructed jet \pT.  The PF jet selections shows a very small rate of unmatched jets at 40 $GeV/c$, significantly lower than Calo jets at the same \pT, which is at least partially driven by the smaller value of $R$.

\begin{figure}[th!]
\begin{center}
\includegraphics[width=0.82\textwidth]{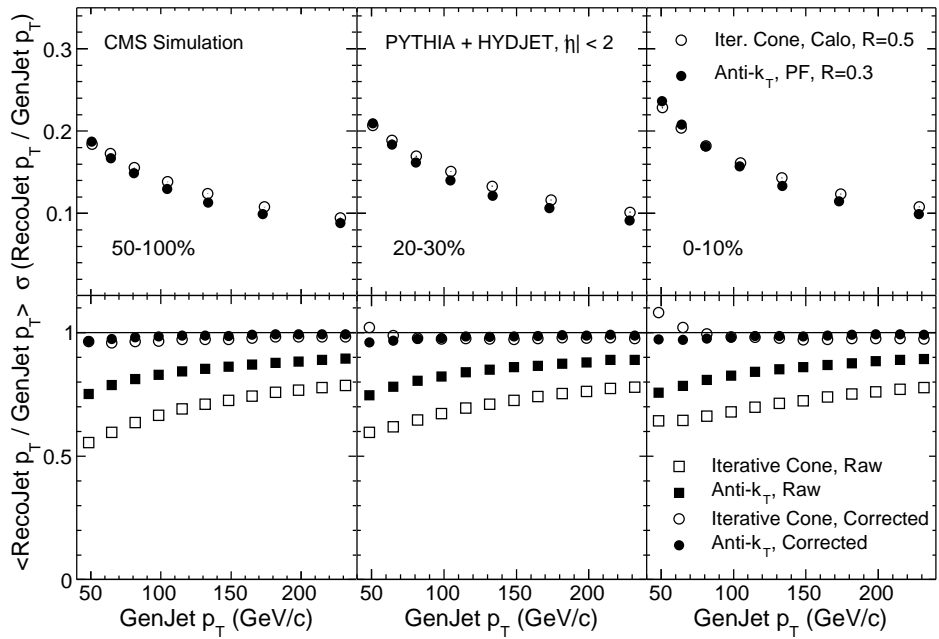}
\caption{The inclusive jet response and resolution for PF and Calo jets for raw (uncorrected -- open symbols) and corrected (closed symbols) jets.}
\label{fig:res}
\end{center}
\end{figure}

Figure 2 further demonstrates the performance of PF jets in {\sc{pythia+hydjet}}, comparing the reconstructed jet \pT\ to that obtained by clustering generator level particles.  The top row shows the jet \pT\ resolution, extracted using a Gaussian fit, for several different centrality selections.  PF  reconstruction improves the jet resolution over most of the \pT\ range. 
This improvement is, however, partially offset by the decreased value of $R$, which increases the degree to which energy can migrate into and out of the jet, when comparing the reconstructed and generator-level jets.   The bottom row of Fig.~\ref{fig:res} shows the jet response, defined as the mean of the ratio of the reconstructed to generated jet \pT, both before and after the application of JECs.  As expected, the raw response to PF jets is closer unity, showing that the size of the required JECs 
is reduced by about half.  The corrected response nicely closes at unity independent of centrality.  This demonstrates the validity of the combined background subtraction + JEC procedure and shows that the particle-based jet reconstruction is insensitive to occupancy effects.

As the JECs corrections are based on an inclusive jet selection in {\sc{pythia}}, the true response may differ from that obtained from simulation.  This may arise if the fragmentation of jets is not accurately modeled, for example if the fraction of quark and gluon jets is different in simulation and data.  This may be of particular concern in heavy-ion collisions where quenching is expected to be larger for gluons than for quarks, and the parton distribution functions are less well known than for protons.  In Fig 3 the corrected response is shown differentially for quark and gluon jets.  One observes a splitting between the quark jets, which tend to fragment harder than average, and gluon jets, which fragment softer.  However, this splitting smaller for PF jets than Calo jets, indicating that PF jet reconstruction is less sensitive to the details of the fragmentation.

\begin{figure}[th!]
\begin{center}
\includegraphics[width=0.51\textwidth]{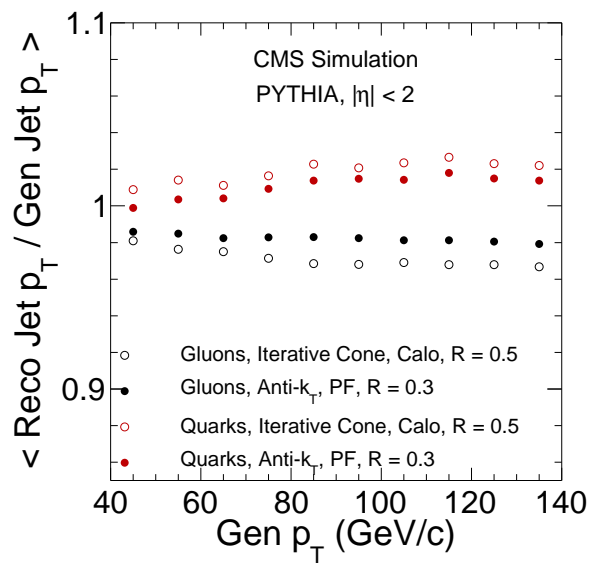}
\caption{The corrected jet response for quark and gluon jets with PF and Calo jet reconstruction in {\sc{pythia}}.}
\label{fig:flavor}
\end{center}
\end{figure}

To conclude, PF reconstruction has been performed in heavy-ion collisions for the first time.  The jet finding efficiency and the dijet mismatch rate have been evaluated in simulated data.  The performance of PF-based jet reconstruction has been demonstrated both in terms of both response and resolution and compared to Calo-based jet finding.   Finally, by studying the response of quark and gluon jets, PF jets were shown to be less sensitive to the details of jet fragmentation than Calo jets.


\section*{References}
\begin{thebibliography}{99}
\bibitem{cms}
CMS Collaboration 2008 {\it JINST} {\bf 3} S08004
\bibitem{PFpp} CMS Collaboration 2009 CMS Physics Analysis Summary CMS-PAS-PFT-09-001
\bibitem{Kodolova}
Kodolova O et. al 2007 {\it Eur. Phys. J. }{\bf 50} 117
\bibitem{ourPaper}
 CMS Collaboration 2011 
  {\it Preprint} arXiv:1102.1957v2
\bibitem{hydjet}
Lokhtin I P and Snigirev A M 2006 {\it Eur. Phys. J. C}{\bf 45} 211
\end{thebibliography}
\end{document}